\documentclass[12pt]{article}
\usepackage{epsfig}
\begin{document}
\newcommand{\be}{\begin{equation}}
\newcommand{\ee}{\end{equation}}

\newcommand{\cc}{\cite}
\newcommand{\ba}{\begin{eqnarray}}
\newcommand{\ea}{\end{eqnarray}}
\begin{titlepage}
\begin{center}

  \vspace{2cm}

  {\Large \bf Role of anomalous chromomagnetic interaction  in
    Pomeron and Odderon  structures  and  in  gluon distribution}
  \vspace{0.50cm}\\
  Nikolai Kochelev\footnote{kochelev@theor.jinr.ru}\\
   { \it Bogoliubov Laboratory of Theoretical Physics, Joint
Institute for Nuclear Research, Dubna, Moscow region, 141980
Russia} \vskip 1ex
\end{center}
\vskip 0.5cm \centerline{\bf Abstract} We calculate the
contribution arising from nonperturbative quark-gluon
chromomagnetic interaction to the high energy total quark-quark
cross section and to  gluon distributions in nucleon. The
estimation obtained within the instanton model of QCD vacuum leads
to the conclusion that this type of interaction gives the
dominating contribution to the Pomeron coupling  with  the light
quarks and to  gluon distribution in  light hadrons at small
virtualities of quarks and gluons. We argue that the Odderon,
which is the $P=C=-1$ partner of Pomeron,
  is governed by the spin-flip
component related to nonperturbative three-gluon exchange induced by
anomalous quark-gluon chromomagnetic interaction.

 \vspace{1cm}
\end{titlepage}
\setcounter{footnote}{0}

\section{Introduction}

The gluon distribution in nucleon is one of the central quantities
in particle physics which determines the  high energy cross
section values of the huge amount of important processes. In spite
of the tremendous achievements   in  the last years  in the
measurement of this distribution, full understanding of the
dynamics of gluons inside hadrons is absent so far (see reviews
\cite{Landshoff:2009wt,Ivanov:2004ax}). In the Regge theory the
behaviour of the  gluon distribution function at small Bjorken $x$
is controlled  by the contribution coming from the  Pomeron
exchange which may have  so-called "soft" and "hard" parts
\cite{Landshoff:2002zy}. Usually, the hard Pomeron is associated
with the perturbative BFKL regime \cite{BFKL} and the soft part is
assumed to be originated from nonperturbative QCD dynamics
\cite{LN}. Nonperturbative effects  arise  from the complex
structure of QCD vacuum. The instantons are one of the well
studied  topological fluctuations of vacuum gluon fields which
might be responsible for many nonperturbative phenomena observed
in particle physics (see reviews \cite{shuryak,diakonov}). Their
possible importance in the structure of the Pomeron and gluon
distribution was considered in quite different approaches
\cite{Kochelev:1997qq}, \cite{shuryakzahed}, \cite{levin},
\cite{Dorokhov:2004fb},\cite{diakonov}
 for the different approximations
to the complicated quark-gluon dynamics in instanton vacuum.
 In particular,  it was shown  \cite{kochelev1}  that
instantons lead to the appearance of
  anomalous  chromomagnetic quark-gluon  interaction (ACQGI).
   It was demonstrated  that
this  new type of quark-gluon  interaction   might be responsible
for the observed large single-spin asymmetries in various high
energy reactions \cite{kochelev1,Cherednikov:2006zn}. Furthermore,
 it gives a large contribution to
the high energy quark-quark scattering cross section
\cite{kochelev2}. The first  estimation of the effect of ACQGI on
nucleon gluon distribution was made in \cite{Kochelev:1997qq} and
small x behavior $g(x)\propto 1/x$ corresponding to soft Pomeron
was found. It was clear from that study that anomalous
chromomagnetic interaction should also play an important role in
the structure of Pomeron. Indeed, recently the  model for soft
Pomeron based on this interaction has been suggested
\cite{diakonov}.

In this paper, we  consider  the detailed structure of the Pomeron
and gluon distribution with the special attention to the interplay
between their perturbative and nonperturbative components.  We
also  discuss the possible manifestation of  ACQGI in  Odderon
exchange.

\section{ Anomalous chromomagnetic  quark-gluon interaction}

In the general case, the interaction vertex of  massive quark with
gluon can be written in the following form:
\begin{equation}
V_\mu(k_1^2,k_2^2,q^2)t^a = -g_st^a[ \gamma_\mu F_1(k_1^2,k_2^2,q^2) -
\frac{\sigma_{\mu\nu}q_\nu}{2M_q}F_2(k_1^2,k_2^2,q^2)],
 \label{vertex}
 \end{equation}
\begin{figure}[htb]
\vspace*{-0.0cm}
\centerline{\epsfig{file=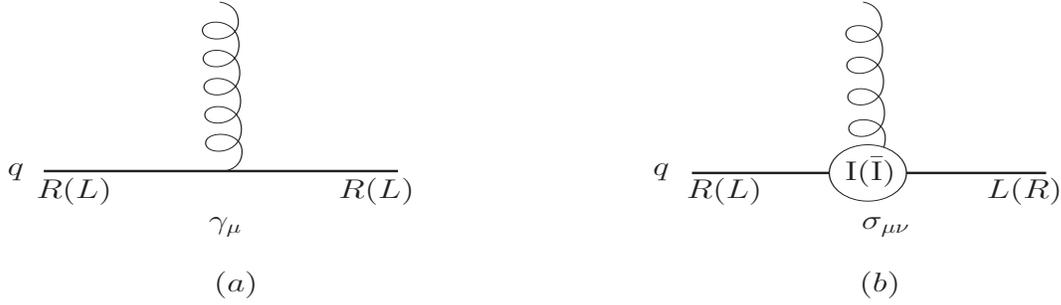,width=4cm,height=14cm,angle=90}}\
\caption{ The quark-gluon coupling: a) perturbative and b)
nonperturbative. Symbols $R$ and $L$ denote quark chirality and symbol $I (\bar I)$ denotes
  instanton (antiinstanton).}
\end{figure}
where the form factors $F_{1,2}$ describe nonlocality of the
interaction, $k_{1,2}$ is the momentum of incoming and outgoing
quarks, respectively, and $ q=k_1-k_2$, $M_q$ is the quark
mass, and $\sigma_{\mu\nu}=(\gamma_\mu \gamma_\nu-\gamma_\nu
\gamma_\mu)/2$. In various applications to high energy reactions
based on perturbative QCD (pQCD) it is usually assumed that only non-spin
flip first term in Eq.(\ref{vertex}) (Fig.1a) contributes and  one can neglect the
second term in  this equation,Fig.1b, because in the limit of the
massless quark this term should be absent due to quark chirality
conservation in massless pQCD. However,  it has recently been shown
that such assumption has no justification in nonperturbative QCD
and second term might give in many cases even a dominant
contribution to  high energy reactions in comparison with the
first one \cite{kochelev1,kochelev2}.

 The  cornerstone of this
phenomenon is the spontaneous chiral symmetry breaking (SCSB)  due
to the complex topological structure of the QCD vacuum. Indeed, the instanton
liquid model for QCD vacuum \cite{shuryak,diakonov} provides
 the  mechanism for such breaking. That mechanism
 is related to the existence of  quark zero modes in the instanton
 field. As the result of SCSB, the
 light quarks in nonperturbative QCD vacuum have  the dynamical
 mass, $M_q$. Additionally, t'Hooft  quark-quark interaction induced by
 quark-zero modes leads to the  violation of $U(1)_A$ symmetry in
 strong interaction.

In high energy  reactions one might naively expect the smallness
of SCSB effects because  of the energy $\sqrt{s}\gg M_q$. Indeed,  it
might be correct for the reactions where the dominating
contribution comes from quark-exchange diagrams. Within the instanton
model this type of diagrams is originated from the t'Hooft quark-quark
interaction contribution. However, instantons also lead to
specific {\it quark-gluon chromomagnetic interaction}
\cite{kochelev1} which is  presented by the second term in
Eq.(\ref{vertex}) (Fig.1b).
It is evident that this term should lead
to a nonvanishing contribution to high energy reactions because
it induces t-channel nonperturbative gluon exchange. The size of
the contribution is determined by the value of  anomalous quark
chromomagnetic moment (AQCM) \footnote{ The  definition of
anomalous AQCM used in Eq.(2) differs
by a factor of two from the corresponding quantity presented in
\cite{kochelev1} and \cite{diakonov}.}
\begin{equation}
\mu_a=F_2(0,0,0). \label{AQCM}
\end{equation}
We should point out that within the instanton model  the shape  of
form factor $F_2(k_1^2,k_2^2,q^2)$ is fixed:
\be
 F_2(k_1^2,k_2^2,q^2) =\mu_a
\Phi_q(\mid k_1\mid\rho/2)\Phi_q(\mid k_2\mid\rho/2)F_g(\mid q\mid\rho) \ , \label{form1} \ee
where
\ba
\Phi_q(z)&=&-z\frac{d}{dz}(I_0(z)K_0(z)-I_1(z)K_1(z)), \nonumber\\
F_g(z)&=&\frac{4}{z^2}-2K_2(z) \label{form3} \ea are the
Fourier-transformed quark zero-mode and instanton fields,
respectively, and $I_{\nu}(z)$, $K_{\nu}(z)$, are the
modified Bessel functions and $\rho$ is the instanton size.

The value of AQCM  is determined by the effective density of the instantons
$n(\rho)$ in nonperturbative QCD vacuum \cite{kochelev1}: \be
\mu_a=-\pi^3\int \frac{d\rho n(\rho)\rho^4}{\alpha_s(\rho)}.
\label{mu}
\ee
 The shape of instanton density in the form
\be
n(\rho)=n_c\delta(\rho-\rho_c),
\label{den}
\ee
leads to
 AQCM  which is proportional to the  packing fraction of instantons  $f=\pi^2n_c\rho_c^4$ in vacuum
\be
\mu_a=-\frac{\pi f}{\alpha_s(\rho_c)}.
\label{mc}
\ee
By using the following relation between parameters of the instanton model \cite{DKE}:
\be
f=\frac{3}{4}(M_q\rho_c)^2,
\label{rel}
\ee
we obtain
\be
\mu_a=-\frac{3\pi (M_q\rho_c)^2}{4\alpha_s(\rho_c)}.
\label{AQCM1}
\ee
This formula coincides with the result for AQCM presented in Eq.(7.2)  in the paper by Diakonov
 \cite{diakonov} and
shows the direct connection between  AQCM and SCSB phenomena.
The dimensionless parameter $ \delta=(M_q\rho_c)^2$ is one of the main parameters of the
instanton model. It is proportional to the packing fraction of instantons in QCD vacuum $\delta\propto f\ll 1$,
Eq.(\ref{rel}), and  is  rather small. For a fixed value of average instanton size $\rho_c^{-1}=0.6$ GeV
 it  changes from
 $\delta^{MF}=0.08$ for $M_q=170 MeV$
in the mean field approximation \cite{shuryak} to $\delta^{DP}=0.33$
 for $M_q=345$ MeV  within Diakonov-Petrov model (DP) \cite{DP}.
For the strong coupling constant at the scale  of instanton average  size \cite{shuryak},\cite{diakonov}
\be
\alpha_s(\rho_c)\approx 0.5,
\label{coupling}
\ee
we  obtain the following values for AQCM:
\be
\mu_a^{MF}\approx-0.4, \ \ \ \mu_a^{DP}\approx-1.6
\label{estim}
\ee
in the mean field approximation and in the  DP approach, respectively.
We would like to emphasize that
in spite of the strong dependence of  AQCM   on the value of the
effective quark mass in QCD vacuum, AQCM is very large in the wide interval the possible
changing of instanton model parameters.
The origin of this peculiarity is in the large numerical factor in front of $\delta$ in Eq.(\ref{AQCM1})
for AQCM.
Indeed,  this formula can be rewritten
in the following form:
\be
\mu_a=-\frac{3}{8}S_0\delta,
\label{ff}
\ee
where $S_0=2\pi/\alpha_s(\rho)$ is the Euclidean instanton action.
The typical value of this action is very large \cite{shuryak,diakonov}
\be
S_0\approx 10\div 15
\label{action}
\ee
and leads to the compensation of  the $\delta$ smallness effect on AQCM.

Within the instanton model approach the first term in Eq.(\ref{vertex})
 is related to the nonzero mode contribution to quark propagator in the instanton field. The nonzero
modes contribution to quark propagator can be approximated with
high accuracy by   perturbative  propagator  \cite{shuryak}.
Due to zero mode dominance  for the light quarks,  \cite{shuryak}, we can expect that for the light
quarks  this sort of contribution should be suppressed in  comparison with the second term in Eq.(\ref{vertex}).
However, for heavy quark the first term should dominate because there are no zero modes for heavy
quark in the instanton field. Furthermore, instanton induced form factors in the chromomagnetic part
of interaction suppress the contribution of the second term for highly virtual quark and/or gluon.
Therefore, form factor in the first term in Eq.(\ref{vertex}) might be chosen in the form
\be
F_1(k_1^2,k_2^2,q^2)=\Theta(\mid k_1^2\mid-\mu^2)\Theta(\mid k_2^2\mid-\mu^2)
\Theta(\mid q^2\mid-\mu^2),
\label{ff2}
\ee
where $\mu$ is the factorization scale between perturbative and nonperturbative regimes.
In our estimation below
 we will use $\mu\approx 1/\rho_c\approx 0.6 $ GeV.

\section{ Fine Pomeron structure}

Let us estimate the contribution of the vertex, Eq.(\ref{vertex}),
to the total high energy quark-quark scattering cross
 section.
The leading diagrams  contributing to the non-spin flip amplitude
of $q-q$ scattering are shown in Fig.2 and for colorless t-channel
exchange presents the model of the Pomeron. The imaginary part of
the total forward scattering amplitude gives the total quark-quark
cross-section.
\begin{figure}[h] \vspace*{-0.0cm}
\centerline{\epsfig{file=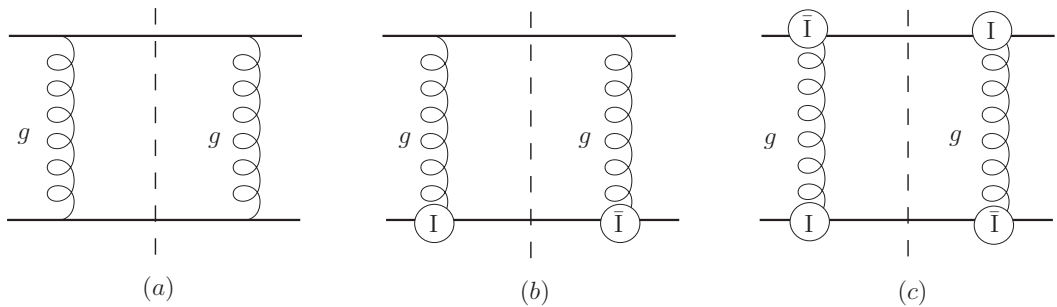,width=4cm,height=14cm,angle=90}}\
\caption{ The fine Pomeron structure in the model with
perturbative interaction and nonperturbative  ACQGI:
 a) perturbative contribution, b)
interference perturbative and nonperturbative vertices, c)
nonperturbative contribution. The symbol $I$ ($\bar I$) denotes instanton (antiinstanton).}
\end{figure}

So, in our model Pomeron includes the pure perturbative exchange
(Fig.2a), nonperturbative (Fig.2c)
 diagrams and the
mixed graph (Fig.2b).

By using the relation, Eq.(\ref{AQCM1}), the total contribution to quark-quark cross section
for the quarks with small virtualities is
\be
\sigma^{total}=\sigma^{pert}+\sigma^{mix}+\sigma^{nonpert},
\label{total}
\ee
where
\be
\sigma^{i}=\int_{{q^2_{min}}}^\infty \frac{d\sigma^i(t)}{dt}dq^2,
\label{ii}
\ee
\ba
\frac{d\sigma(t)^{pert}}{dt}&=&\frac{8\pi\alpha_s^2(q^2)}{9q^4}\nonumber\\
 \frac{d\sigma(t)^{mix}}{dt}&=&\frac{\alpha_s(q^2)\pi^2\mid\mu_a\mid\rho_c^2F_g^2(|q|\rho_c)}{3q^2}\nonumber\\
 \frac{d\sigma(t)^{nonpert}}{dt}&=&\frac{\pi^3\mu_a^2\rho_c^4F_g^4(|q|\rho_c)}{32},
\label{cross}
\ea
where $q^2=-t$ and $q^2_{min}\approx 1/\rho_c^2$ for perturbative and mixed contributions and
$q^2_{min}=0$ for pure nonperturbative (Fig.2c) contribution.

\begin{figure}[h] \vspace*{-0.0cm}
\centerline{\epsfig{file=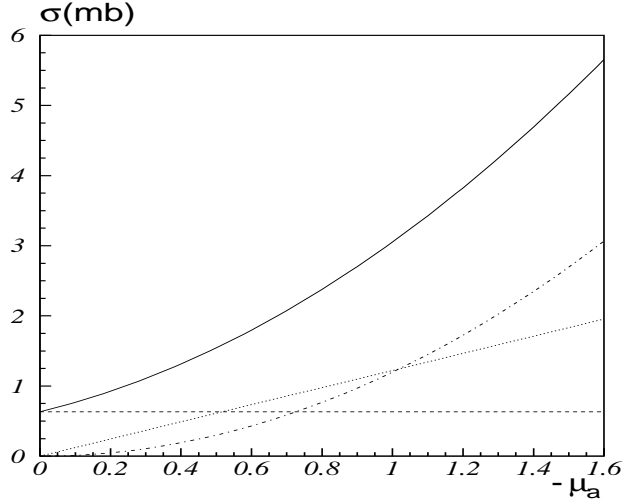,width=10cm,height=7cm,angle=0}}\
\caption{ The contibution to the total quark-quark cross section as a function of AQCM:
 perturbative (dashed line) ,  mixed (dotted line), nonperturbative (dashed-dotted line) and their sum (solid line).}
\end{figure}

For the strong coupling constant, the following parametrization
was used for the case $N_f=3$:
\begin{equation}
\alpha_s(q^2)=\frac{4\pi}{9\ln((q^2+m_g^2)/\Lambda_{QCD}^2)},
\end{equation}
where $\Lambda_{QCD}=0.280$ GeV  and the value $m_g=0.88$ GeV
 was fixed
from the requirement $\alpha_s(q^2=1/\rho_c^2)\approx \pi/6$ \cite{diakonov}.
This form describes the frozen coupling constant in the infrared region, $\alpha_s(q^2)\rightarrow constant$ as
 $q^2\rightarrow 0$.

The result of  calculation of the different contributions to the total quark-quark scattering
cross section is presented in Fig.3 as a function of  AQCM. It is evident that within
the interval $0.4<\mid\mu_a\mid<1.6$ the main contribution  comes from the terms related to the
anomalous quark-gluon chromomagnetic interaction.
Recently, the effects of nonzero AQCM in hadron spectroscopy  has been considered (see
\cite{Ebert:2009ub} and references therein). It was shown that the value of
\be
\mu_a=-1
\label{muaf}
\ee
 is favoring to describe
the fine structure of hadron spectrum. This value of AQCM  corresponds to dynamical quark mass $M_q=280$ MeV.
This mass is in  agreement with recent result of  analysis
of dressed-quark propagator within DSE approach involving  the lattice-QCD data from
 \cite{Bhagwat:2006tu}.
 We will adopt this value
 in our estimations below.
For that set of  parameters the  total quark-quark cross section $\sigma_{qq}^{total}=3.05 $ mb
 is the sum of the following  partial cross sections:
\be
\sigma_{qq}^{pert}=0.63 mb,\ \ \sigma_{qq}^{mix}=1.22 mb, \ \  \sigma_{qq}^{nonpert}=1.21mb,
\label{cross2}
\ee
and it is  not far away from "experimental" quark constituent
 model value $\sigma_{qq}^{exp}\approx 4$ mb, which is needed to describe the inelastic proton-proton
and proton-antiproton cross-sections; $\sigma_{PP(\bar
P)}^{in}=36$ mb in the energy range where they are approximately
constant. One may expect also  an additional contribution to the
total cross section arises   from the multigluon and multiquark
emission induced by the quark-gluon-instanton  vertex. It will
bring our esimation to the experimental value. It follows from
Eq.(\ref{cross2}) that the  contribution to the quark-quark cross
section due to non-perturbative chromagnetic quark-gluon
interaction  is about $80\% $ and the contribution from pure
perturbative exchange is about $20\%$ and  quite small. Therefore,
within our model the dynamics of soft Pomeron is determined
 not by the  $\gamma_\mu$-like quark-gluon vertex (Fig.1a) as in most conventional models for the Pomeron, but
by the $\sigma_{\mu\nu}$ vertex pictured in Fig.1b. The widely assuming statement is that the
 difference in the dynamics of soft and hard
Pomerons   comes from  the difference in their dependence on such
kinematic variables as  total energy and transfer momenta. From
our point of view, the main source of  difference between two
exchanges arises   from a completely different spin structure of
quark-gluon interaction inside the Pomeron exchange.

In our above estimation above only  simplest contributions to the
Pomeron exchange presented in Fig.2 was considered. Due to pure
spin one t-channel exchange they lead to cross section independent
of the energy. Therefore the effective Pomeron intercept
$\alpha_P=1$ in this approximation. It is well known that the
experimental data show that the value of soft Pomeron intercept
$\alpha_P(0)\approx 1.08$ \cite{Donnachie:1992ny}. In spite of the
fact that empirically  soft Pomeron intercept  close to one, its
deviation from one leads to visible energy dependence of the total
and  diffractive cross sections and to a  large subleading
contributions at very high energies.
\begin{figure}[h] \vspace*{-0.0cm}
\centerline{\epsfig{file=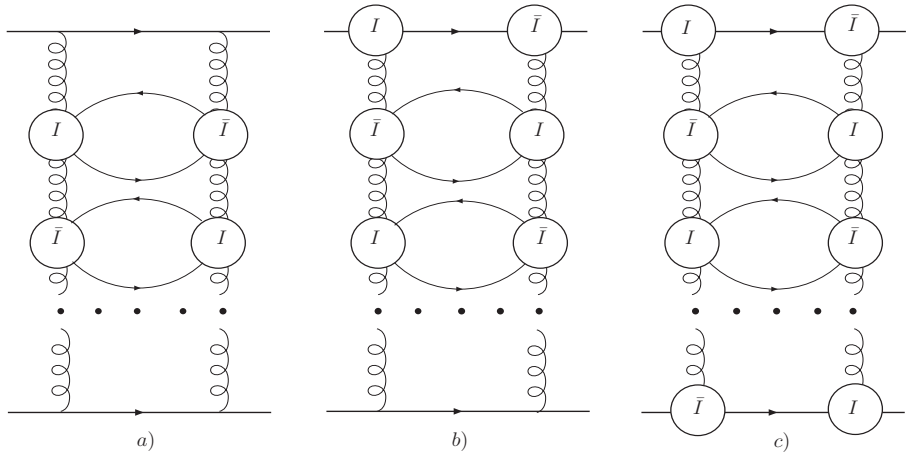,width=6cm,height=12cm,angle=90}}\
\caption{ The example of the diagrams which give  the contribution
to energy-dependent part of  Pomeron exchange.}
\end{figure}
Some of diagrams which  provide such subleading contributions in
our model   are presented in Fig.4. It is evident that at low
energy such contributions should be suppressed by even powers of
small packing fraction of instantons in QCD vacuum,$f^n<1/4^n,
n=2,4...$. However, due to their logarithmic growth with
increasing of energy they might give the dominant contribution at
very large energy. The
 calculation of
these contributions is beyond of this paper and will be the
subject of the separate publication \cite{kochelevfuture}.

\section{Chromomagnetic Odderon}

Within the conventional approach, the Odderon P=C=-1 partner of
Pomeron, originates from  three gluon exchange (Fig.4a) with
non-spin-flip perturbative-like quark-gluon vertex
\cite{Gauron:1986nk},
\cite{Donnachie:1990wd},\cite{Bartels:1999yt},
\cite{Ewerz:2003xi},\cite{Braun:2007kz}.
  The experimental support of the existence  of such exchange comes
from high energy ISR data on the difference in the dip structure
around $\mid t\mid \approx 1.4 $ GeV between the proton-proton and proton-antiproton
differential cross sections \cite{Breakstone:1985pe}.
 However, there is no any signal for the Odderon at very small transfer
momentum $t$ \cite{Landshoff:1998fp}.

\begin{figure}[h] \vspace*{-0.0cm}
\centerline{\epsfig{file=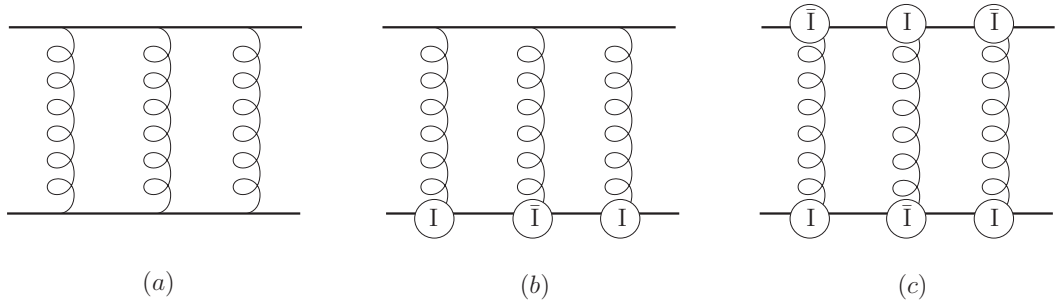,width=4cm,height=14cm,angle=90}}\
\caption{ The structure of Odderon exchange: a) non spin-flip
perturbative three gluon exchange, b) and c) nonperturbative
spin-flip contributions.}
\end{figure}
According to our model, the perturbative part of the Odderon,
Fig.5a, in the region of momentum transfer $\mid t\mid/9\leq
1/\rho_c^2$  is expected to be much smaller in comparison with the
nonperturbative part presented by the graphs, Fig.5b and Fig.5c
\footnote{ The detailed calculation will be published elsewhere.}.
\begin{figure}[h] \vspace*{-0.0cm}
\centerline{\epsfig{file=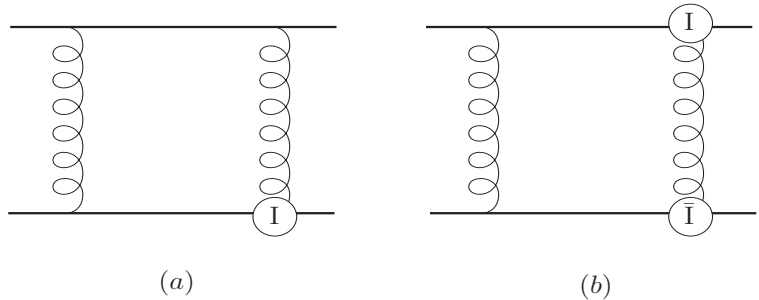,width=4cm,height=10cm,angle=90}}\
\caption{ The example of the diagrams which give the contribution
to spin-flip component of Pomeron.}
\end{figure}

It is clear that the first diagram gives  rise  to the  non
spin-flip amplitude of quark-quark scattering, the diagram in
Fig.5b leads to  single spin-flip and the diagram in Fig.6c
presents double spin flip (see helicity structure of vertices in
Fig.1).
 By using the conventional  notation for helicity amplitudes
$\Phi_n=<\lambda_{i_1}\lambda_{i_2}\mid \lambda_{f_1}\lambda_{f_2}>$ (see e.g.
\cite{Bourrely:1980mr}), where $n=1,...,5$ and
 $\lambda_{i_{1,2}(f_{1,2})}$ are
helicities of initial (final) quarks, respectively, one can see
that the graph in Fig.5a gives the contribution to
 the $\Phi_1$ and $\Phi_3$ amplitudes, diagram in Fig.5b contributes to the  $\Phi_5$ amplitude, and Fig.5c
gives  rise to the $\Phi_4$ amplitude. Our conjecture is that  the
spin-flip amplitude dominates  in  Odderon exchange. Therefore,
one might expect that  Odderon should strongly interfere  with the
spin-flip part of Pomeron. Some of the diagrams which give the
rise  to the spin-flip part of the  Pomeron are presented in
Fig.6.

 We would like to mention that in
\cite{Zakharov:1989bh},
 \cite{Kopeliovich:1989hp} and \cite{goloskokov} an
 alternative mechanism for the spin-flip component of Pomeron and
 Odderon \cite{Zakharov:1989bh}, was discussed. This mechanism is based on
 the  existence of the quark-diquark component in the
 nucleon wave function.

\section{Gluon distribution and chromomagnetic quark-gluon interaction}

It was shown above that the Pomeron structure is rather
complicated. It includes  perturbative, "hard", and
nonperturbative, "soft", parts and their interference, "soft-hard"
part.

\begin{figure}[h]
\hspace*{0.5cm} \vspace*{0.5cm}
 \centering
\centerline{\epsfig{file=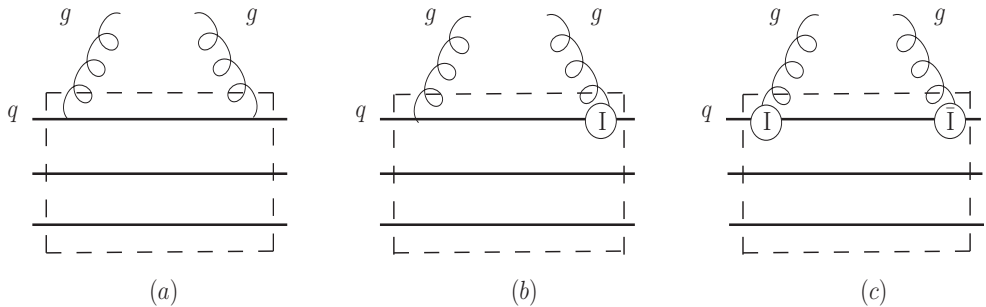,width=4cm,height=13cm,angle=90}}\
\caption{The diagrams contributing to nucleon gluon distribution: a) "hard"-perturbative,
b) "hard-soft" interference perturbative and nonperturbative exchanges, c) "soft"-nonperturbative
part.}
\end{figure}

Therefore, this structure  should also manifest itself in the
structure of gluon distribution in nucleon. One of the ways to
show it is in the use of a DGLAP-like approach
\cite{DGLAP},\cite{AP} with the modified quark splitting function
${\cal{P}}_{Gq}$ according with the vertex, Eq.(\ref{vertex})
\cite{Kochelev:1997qq}. The diagrams giving the contribution to
nucleon gluon distribution  in our model are presented in Fig.7.

\begin{figure}[h]
\begin{minipage}[c]{8cm}

\vskip -1.5cm
\hspace*{0.5cm} \epsfig{file=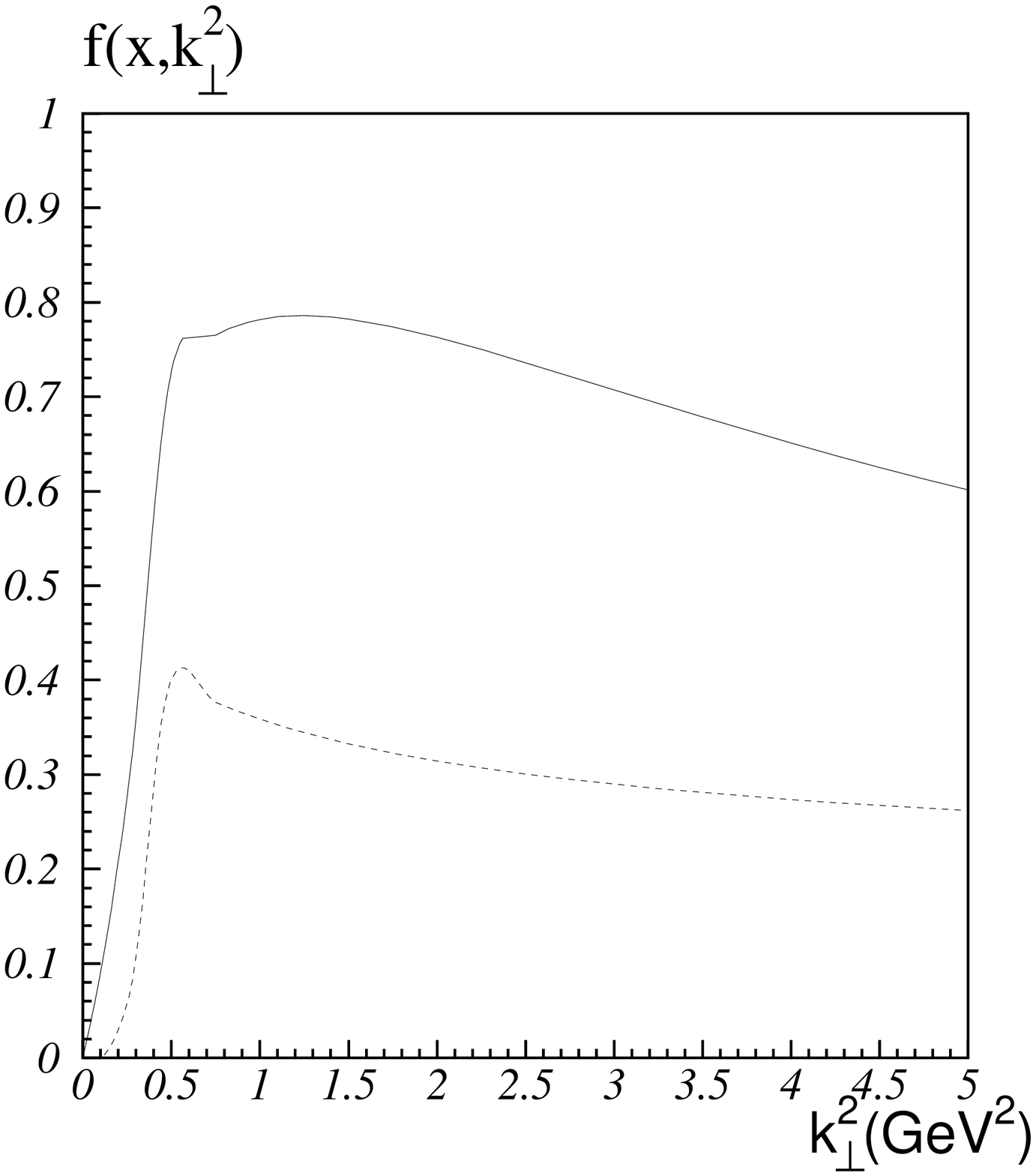,width=7cm,height=5cm}
\caption{The unintegrated gluon distribution at $x=10^{-2}$: solid (dashed) line is total (perturbative) contribution. }
\end{minipage}
\begin{minipage}[c]{8cm}
\hspace*{0.5cm}
\epsfig{file=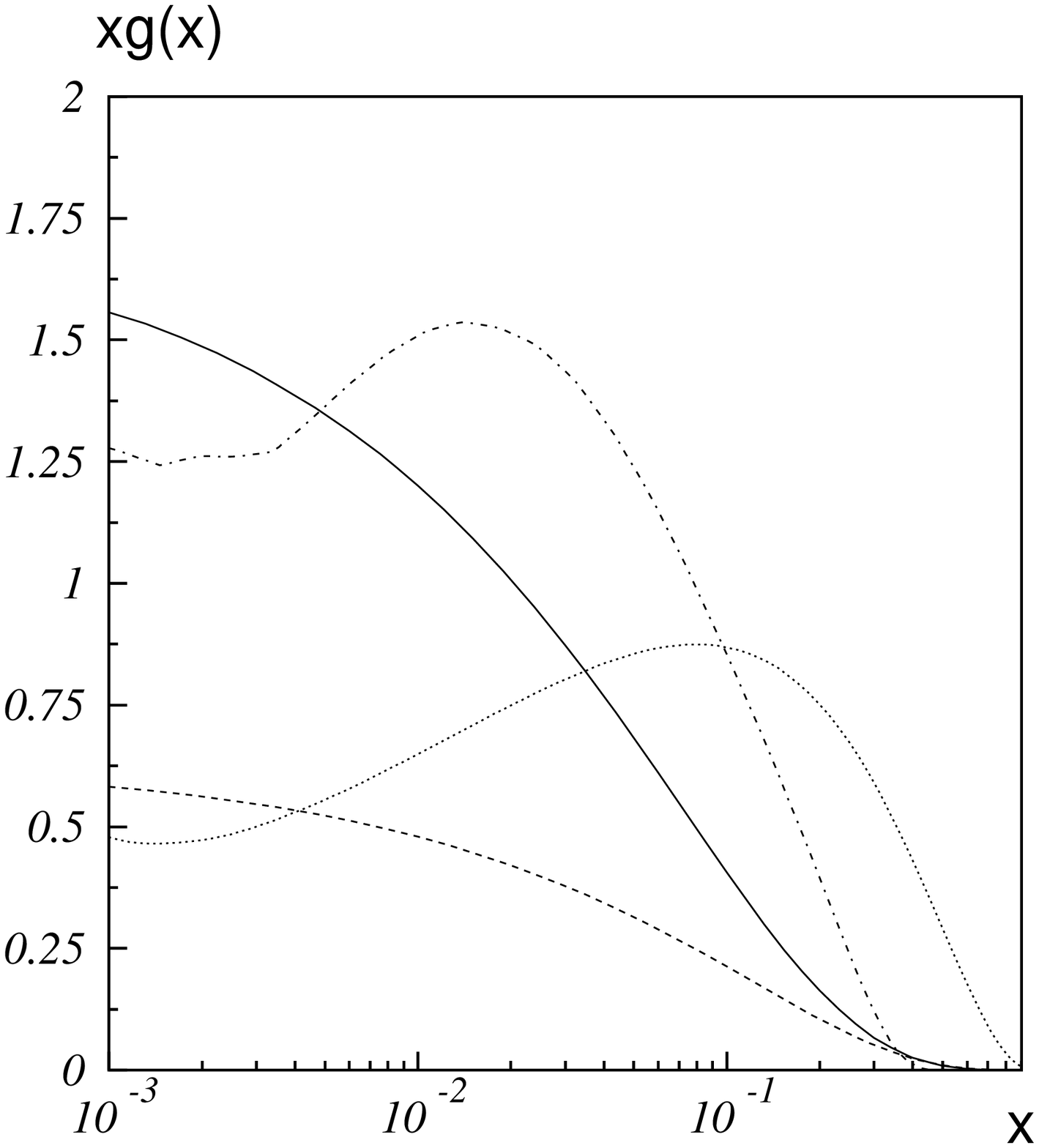,width=7cm,height=5cm, angle=0}
\caption{Perturbative  (dashed line)  and total (solid
line) contributions to gluon distribution at $Q^2=1GeV^2$ in comparison with some of the phenomenological fits:
dotted line is ALEKHIN02LO set and dashed-dotted line is MSTW2008LO fit \cite{PDF}.}

\end{minipage}
\end{figure}

At  present, unintegrated gluon distibution
is widely used in  different applications  (see, for example, \cite{Kimber:2001sc,Ivanov:2000cm}).
To  calculate this distribution, we  use the convolution model formula
\be
f(x,k^2_\bot)=N_qk^2_\bot\int_x^1\frac{dy}{y}{\cal{P}}_{Gq}({x}/{y},k^2_\bot)q_V(y),
\label{un}
\ee
where $N_q=3$ is the number of valence quarks in nucleon,
$q_V$ is the valence quark distribution function in nucleon,  ${\cal{P}}_{Gq}$ is the quark splitting
function as defined in \cite{AP}, and we neglect possible intrinsic momentum  dependence of $q_V$ related to the confinement scale.

The splitting function for the general vertex Eq.(\ref{vertex}) is given by the formula
\be
 {\cal{P}}_{Gq}(z,k^2_\bot)=\frac{C_Fz(1-z)k^2_\bot}{8\pi^2(k^2_\bot+M_q^2z^2)^2}
\sum_{\lambda}Tr\left\{(\hat k_C+M_q) U_\mu(t)(\hat k_A+M_q)\bar U_\rho(t)\right\}\epsilon_\mu(\lambda)\epsilon_\rho^*(\lambda),
\ee
where $U_\mu(t)=V_\mu(0,0,t)$, $k_A$ $(k_C)$ is momentum of initial (final) quark, $t=q^2=(k_A-k_C)^2$,
$\bar U=\gamma_0U^\dagger\gamma_0$ and $\lambda$ is gluon helicity. In the infinite momentum frame
\ba
k_A&=&(P,P+\frac{M_q^2}{2P},\vec 0_\bot)\nonumber\\
k_C&=&((1-z)P+\frac{k^2_\bot+M_q^2}{2(1-z)P},(1-z)P,-\vec k_\bot)\nonumber\\
q&=&(zP-\frac{k^2_\bot+M_q^2z}{2(1-z)P},zP,\vec k_\bot),
\label{kin}
\ea
the result for splitting function is
\ba
{\cal{P}}_{Gq}(z,k^2_\bot)&=&\frac{C_Fk^2_\bot}{2\pi z(k^2_\bot+M_q^2z^2)^2}\nonumber\\ &\times &[
                  (\sqrt{\alpha_s(\mid t\mid)}\Theta(\mid t\mid-\Lambda^2) +\sqrt{ \alpha_s(1/\rho_c^2)}\mu_a
                  F_g(\mid t\mid))^2z^2 \nonumber\\&+&
        2((1 - z)\alpha_s(\mid t\mid)\Theta(\mid t\mid-\Lambda^2)+
              \frac{\alpha_s(1/\rho_c^2)\mu_a^2k_\bot^2}{4M_q^2}F^2_g(\mid t\mid))],
\label{split2} \ea where $\mid t\mid=(k_\bot^2+M_q^2z^2)/(1-z)$ is
the gluon virtuality  in Fig.7.

The integrated distribution is given by
\be
g(x,Q^2)=\int_{0}^{Q^2}\frac{dk^2_\bot}{k^2_\bot} f(x,k^2_\bot),
\label{in}
\ee

For estimation  we  use a simple form for valence quark
distribution \be q_V(x)=1.09\frac{(1-x)^3}{\sqrt{x}} \label{val}
\ee with the normalization \be \int_0^1q_V(x)dx=1. \nonumber \ee
The result of calculation of unintegrated gluon distribution at
$x=10^{-2}$ is presented in Fig.8 as a function of $k^2_\bot$. The
result for integrated gluon distibution  at small\\ $Q^2=1$
GeV~$^2$ is pictured in Fig.8. It is evident that the
nonperturbative contribution dominates  in both unintegrated and
integrated gluon distributions. For the large $Q^2$ perturbative
contribution starts to dominate due to its  stronger $Q^2$
dependence. Such a difference in the $Q^2$ dependence is directly
related to the difference in the $k^2_\bot$ behavior between
perturbative and nonperturbative contributions coming from the
spin-non-flip and spin-flip part of the general quark-gluon
vertex, Eq.(\ref{vertex}). In  Fig.9 we also present the
comparison of our result with some available phenomenological
parametrizations. By taken into account  the uncertainties in the
extraction of gluon distribution from the data and our simple
parametrization  for valence quark distribution we may say that
agreement is rather good.

\section{Conclusion}
In summary, we suggest a new approach to the Pomeron and Odderon
structures and gluon distribution in hadrons. It is based on the
modified quark-gluon vertex which includes the non-perturbative
spin-flip part related to anomalous chromomagnetic interaction. It
is shown that this interaction gives the main contribution to the
Pomeron coupling to small virtuality light quarks and to the gluon
distribution in nucleon. Our conjecture is that the origin of the
difference between "soft" and "hard" Pomerons is related to the
difference in the  spin structure of quark-gluon interaction
governing these effective exchanges. We give the arguments in
favor of the spin-flip dominance in Odderon exchange.

\section{Acknowledgments}
The author is very grateful to I.~O.~Cherednikov, A.E. Dorokhov,
S.~V.~Goloskokov,  E.A.~Kuraev, N.N.~Nikolaev and L.N. Lipatov
   for useful
discussions. This work was supported in part   by  RFBR grant 10-02-00368-a, by  Belarus-JINR grant,
and by Heisenberg-Landau program.

\end{document}